\documentclass[aps,prd,final,twocolumn,preprintnumbers,floatfix,nofootinbib,10pt,superscriptaddress]{revtex4-2}

\usepackage{graphicx}
\graphicspath{ {./images/} }

\usepackage{amsmath}
\usepackage{bm}
\usepackage{ocgx2}
\usepackage{mathrsfs}
\usepackage{comment}
\usepackage{lipsum}
\usepackage{amsfonts}
\usepackage{amssymb}
\usepackage{amstext}
\usepackage{helvet}
\usepackage{microtype}
\usepackage{esint}
\usepackage{tikz}
\usepackage{mathtools}
\usepackage{scrextend}  
\usepackage[pdftex]{hyperref}
\usepackage{etoolbox}
\usepackage[export]{adjustbox}
\usepackage{bbm}
\usepackage[caption=false]{subfig}
\usepackage{svgcolor}
\usepackage{mathtools}
\usepackage{bbm}

\usepackage{slashed}                      %
\usepackage{multirow}                     %
\usepackage{rotating}                     %
\usepackage{ifthen}                       %
\usepackage{xspace}                       %
\usepackage{comment}
\newcommand{\sref}[2]{\hyperref[#1]{\ref*{#1}#2}}
\newcommand{\iu}{\mathrm{i}\mkern1mu}

\usetikzlibrary{decorations.pathmorphing,patterns}
\usetikzlibrary{ decorations.markings}
\usetikzlibrary{calc}

\newcommand{\pdv}[3][1]{
    \ifnum#1=1
        \frac{\partial #2}{\partial #3}
    \else
        \frac{\partial^{#1} #2}{\partial #3^{#1}}
    \fi
}

\newcommand{\comma}{{,\ \ \ }}

\DeclarePairedDelimiterX\braket[2]{\langle}{\rangle}{#1 \delimsize\vert #2}

\definecolor{orange}{rgb}{1,0.4,0}
\definecolor{green}{rgb}{0,0.65,0}
\definecolor{rossos}{rgb}{0.8,0.2,0.3}
\definecolor{bluscuro}{rgb}{0.15, 0.2, .85}
\definecolor{bluchiaro}{cmyk}{1,.3,0.,0.1}
\hypersetup{colorlinks, citecolor=bluscuro, linkcolor=bluscuro, urlcolor=bluscuro}

\makeatother   %

\newcommand{\GeV}{{\rm \,GeV}}

\begin{document}
	
	\title{Large-Field Vacuum Decay in General Multi-Scalar Theories}
	
	\author{Giorgio Busoni}
	\email{giorgio.busoni@adelaide.edu.au}
	\affiliation{ARC Centre of Excellence for Dark Matter Particle Physics,Department of Fundamental and Theoretical Physics,Research School of Physics, The Australian National University,Canberra, Australian Capital Territory 2601, Australia}
	\affiliation{ARC Centre of Excellence for Dark Matter Particle Physics, University of Adelaide,
		South Australia 5005, Australia}
    \author{Florian Goertz}
    \email{florian.goertz@mpi-hd.mpg.de}
    \affiliation{Max-Planck-Institut f\"ur Kernphysik, Saupfercheckweg 1, 69117 Heidelberg, Germany}
    
	\author{Navneet Krishnan}
	\email{navneet.krishnan@anu.edu.au}
	\affiliation{ARC Centre of Excellence for Dark Matter Particle Physics,Department of Fundamental and Theoretical Physics,Research School of Physics, The Australian National University,Canberra, Australian Capital Territory 2601, Australia}
	
	\author{Rickson Wielian}
	\email{rickson.wielian@student.unimelb.edu.au}
	\affiliation{ARC Centre of Excellence for Dark Matter Particle Physics,Department of Fundamental and Theoretical Physics,Research School of Physics, The Australian National University,Canberra, Australian Capital Territory 2601, Australia}
    \affiliation{School of Physics, University of Melbourne, Parkville, VIC 3010, Australia}
	\date{\today}
	
	\begin{abstract}
    Many theories beyond the standard model exhibit multiple scalar particles. Such multi-scalar theories can in principle host lower-energy vacua, and thus predict that our universe has a finite lifetime due to false vacuum decay. This scenario cannot be ruled out \emph{a priori} as even the standard model's electroweak vacuum has been shown to be metastable; however, for theoretical consistency, we still require that the model does not predict a lifetime much smaller than the age of the universe. The calculation of these tunneling rates at leading order for multi-scalar theories typically includes numerical approaches, or approximations which are frequently not analytically controlled. In this article we show that, in the large-field regime, a one-dimensional radial bounce always produces the exact dominant contribution to the leading order tunneling rate, with corrections being exponentially suppressed. This allows us to write simple analytical expressions to calculate the tunneling rate in multi-scalar theories, in terms of an effective quartic coupling $\lambda_\text{eff}$. For theories with biquadratic scalar potentials, we also derive straightforward analytical expressions for $\lambda_\text{eff}$ in terms of the original theory's couplings. Finally, we provide example applications of our results to study the vacuum stability of the 2HDM+$a$ model and the 3-3-1 model.
	\end{abstract}
	
	\maketitle

\section{Introduction}
\label{sec:intro}

The Standard Model (SM) of Particle Physics has been remarkably successful in describing and predicting the behaviour of fundamental particles, culminating in the discovery of the Higgs boson at the Large Hadron Collider (LHC) in 2012 \cite{CMS:2012qbp,ATLAS:2012yve}. Despite its successes, it is widely accepted that the SM is incomplete as many physical observations remain unexplained, including the origin of neutrino masses \cite{Super-Kamiokande:1998kpq}, the existence of Dark Matter \cite{Jungman:1995df,Bertone:2016nfn}, and the baryon asymmetry of the universe \cite{Kuzmin:1985mm}. 

These shortcomings strongly motivate extensions of the SM involving new fields and interactions. Well-studied frameworks such as supersymmetry (SUSY) \citep{Nilles:1983ge} and grand unified theories (GUTs) \citep{Georgi:1974sy} provide compelling examples of such extensions, often predicting richer scalar sectors beyond the single Higgs doublet of the SM. In particular, SUSY inherently requires at least two Higgs doublets, as realized in the Minimal Supersymmetric Standard Model (MSSM), and may accommodate even larger scalar sectors. This motivates the consideration of simplified but phenomenologically rich extensions such as two-Higgs-doublet models augmented by additional states, e.g. the 2HDM$+a$ framework~\cite{LHCDarkMatterWorkingGroup:2018ufk,Bauer:2017ota}.
In general, extended scalar sectors are well motivated from a cosmological perspective. In the SM, the electroweak phase transition is a crossover for the observed Higgs boson mass, precluding the possibility of electroweak baryogenesis \citep{Kuzmin:1985mm,Kajantie:1996mn}. By enlarging the scalar sector, however, the finite-temperature Higgs potential can be significantly modified, allowing for a strong first-order electroweak phase transition and even multi-step phase transitions \citep{Vaskonen:2016yiu,Fabian:2020hny} that can realize successful baryogenesis \citep{Profumo:2007wc,Astros:2023gda}. Together with dark matter constructions, this provides an additional, complementary motivation for considering scenarios with extended scalar sectors and a variety of corresponding models have been proposed.

Alongside 2HDM-based (and related) setups, gauge extensions such as 3-3-1 models offer an alternative route to new physics. These theories enlarge the electroweak gauge group and also naturally predict additional scalar multiplets \cite{Pisano:1992bxx,Frampton:1992wt,Singer:1980sw,Valle:1983dk}. Interestingly, 3-3-1 models can be embedded in or motivated by higher-scale unification structures and emerge from certain GUT constructions. Moreover, they can explain the number of fermion generations via anomaly-cancellation conditions. As such, they provide a complementary laboratory to study the interplay between extended gauge and scalar sectors, with implications both for collider phenomenology and cosmology.

Scalar fields are in fact unique, as their invariance under Lorentz transformations allows them to have nonzero vacuum expectation values (vevs). As a result, if the scalar potential exhibits multiple minima, the theory can support multiple vacua, each with its own physical predictions. In such a scenario, if the desired vacuum reproducing experimental observations is not the deepest minimum of the potential, then the fields can tunnel to a lower-energy vacuum configuration.

This would be phenomenologically viable as long as the tunneling time is not too short compared to the age of the universe. We call the vacuum in such scenarios `metastable'. However, if the tunneling rate is too large, the theory is inconsistent, and we must either introduce new physics which stabilizes the vacuum, or modify the theory entirely. Correspondingly, we call the vacuum in such scenarios `unstable'.

Since the first measurements of the Higgs mass, it has been known that the Standard Model (SM) vacuum is metastable due to high-energy radiative corrections to the Higgs potential~\cite{Elias-Miro:2011sqh}. This is a reflection of the fact that these tunneling rate calculations provided some of the most stringent theoretical bounds on the Higgs mass, under the assumption that the SM is valid up to some large scale $\Lambda$. For example, Ref.~\cite{Isidori:2001bm} found a lower bound $m_H>115\GeV$, for a cutoff at the Planck scale $\Lambda \sim 10^{19}\GeV$, which exceeded the experimental lower bound of the time $m_H>113\GeV$. This indicates that the experimental observations of the time did not significantly weaken the SM hypothesis. The stringency of these bounds thus motivates us to study the stability of Beyond the SM (BSM) theories, to constrain their viable parameter space and better understand their phenomenology.

The technique to calculate the vacuum decay rate was first introduced by Coleman and Callan \cite{Coleman:1977py,Callan:1977pt}, and has since been refined by various authors \cite{Andreassen:2016cvx,Lee:1985uv,Isidori:2001bm}. The decay rate per unit volume is given by
	\begin{align}
		\Gamma/L^3 =   A e^{-B}\,.
	\end{align}
	Here, $A$ parametrizes quantum fluctuations, $L^3$ is the spatial volume, and $B$ is the Euclidean action of the bounce, which is an O(4)-symmetric field with radial dependence dictated (at leading order) by the bounce equation \cite{Coleman:1977py}\,,
	\begin{align}\label{eqn:bounceeqn}
		\pdv[2]{\phi_i}{\rho}+\frac{3}{\rho} \pdv{\phi_i}{\rho}=\pdv{V}{\phi_i}\,,
	\end{align}
	with boundary conditions $\bm\phi'(0)=0$ and $\bm\phi(\infty)=\bm\phi_\text{FV}$, the false vacuum. In this article, we will focus on the leading contribution $B$, taking into account only quantum corrections that are resummed via the RG-running of the couplings in the model.   
    
	In case of the SM, the instability is generated at only large field values $ \phi \gg 246\,$GeV, so the potential is well-approximated by a quartic
    \begin{align}
        V \simeq \frac{1}{4}\lambda \phi^4,
    \end{align}
    where due to radiative corrections, $\lambda<0$. With this approximation, an analytical solution to the bounce equation \eqref{eqn:bounceeqn} exists, and the bounce action is simply given by \cite{Fubini:1976jm,Lee:1985uv}
	\begin{align}\label{eqn:onedimensionalbounceaction}
		B=\frac{8\pi^2}{3|\lambda|} \,.
	\end{align}
    
	The requirement that the decay rate is not too large compared to the age of the universe $\Gamma\, t_U \lesssim 1$, where $t_U=10^{10}\,$yr, then translates to a direct bound \cite{Isidori:2001bm}
	\begin{align}\label{eqn:lambdabound}
		\lambda(\mu) > \frac{-0.065}{1+0.01\ln(\mu/v)}
	\end{align}
	that needs to be met for all renormalization scales $\mu$. Here, $v=246\,$GeV is the SM vev. We will call the first scale $\Lambda_I$ at which $\lambda(\Lambda_I)$ violates this bound the instability scale. Since a quantum field theory cannot be valid up to this scale, and still be consistent with our existence, this instability scale provides an estimate for the upper bound for the cutoff of the theory.
    
	The general conclusion of such vacuum stability calculations in a quantum field theory is that a large portion of the parameter space will be theoretically disfavored, as the set of viable completions is reduced only to those theories which stabilize the vacuum. On the other hand if an (effective) quantum field theory is `confirmed' with an unstable vacuum, then it would be very exciting news for particle physics: To stabilize the vacuum, an extension to the model must be verifiable at energies below $\Lambda_I$, which we typically find to be far below the grand unification scale $10^{16}\,$GeV. In addition, such extensions would be greatly constrained by vacuum decay, as the high-energy contributions introduced by the extension must offset the negative contributions to $\lambda(\mu)$ by the low-energy theory.

However, while the vacuum stability of single-scalar theories can always be straightforwardly approximated using equation \eqref{eqn:lambdabound}, up to now, similarly simple criteria for multi-scalar theories are not known. Previous investigations have either integrated the multi-field bounce equation~\eqref{eqn:bounceeqn} numerically \cite{Masoumi:2016wot,Guada:2020xnz}, which is less straightforward due to the awkward boundary conditions, or constrained the fields to lie on a straight line \cite{Aravind:2014pva,Greene:2013ida,Chakrabarty:2016smc}. In general, the true bounce does not lie on a straight line and, without further analysis, a straight line bounce can only be used as an uncontrolled approximation to the true bounce. However, in this article, we {\it show} that in the limit where $\Lambda_I\gg v$, a straight line bounce provides the exact leading-order tunneling action, thereby circumventing the difficulties in integrating the multifield bounce equation~\eqref{eqn:bounceeqn}. The bounce action can thus be written succinctly as
	\begin{align}\label{eqn:bounceactioneff}
		B = \frac{8\pi^2}{3|\lambda_\text{eff}|}\,,
	\end{align}
	where $\lambda_\text{eff}$ can be obtained directly by minimizing the quartic potential in the angular directions.

The paper is structured as follows: In Section \ref{Sec:radialbounce}, we demonstrate that the Fubini instanton, restricted on a radial line, provides the dominant leading order contribution to the multi-field tunneling rate. We then derive, in Section \ref{Sec:tunnelingrates}, straightforward expressions for the tunneling rates when the quartic potential can be written as a biquadratic form. Finally, in Section \ref{Sec:2HDMa}, we apply our technique to study the vacuum stability of the Two-Higgs-Doublet-Plus-Pseudoscalar (2HDM+$a$) model, indeed a popular simplified model for dark matter and collider phenomenology, while in Section~\ref{Sec:3-3-1} we treat the
3-3-1 gauge extension of the SM, which can address various of its shortcomings and features a very rich scalar sector -- both of them introduced above.

\section{Radial Bounce for Multi-Scalar Fields}
\label{Sec:radialbounce}
	
Assuming $\Lambda_I \gg v$, we follow the same procedure as in the SM, in the $\overline{\text{MS}}$ scheme, and assume that the quartic part of the potential dominates. Denoting the quartic part of the potential by $V^{\{4\}}(\bm\phi)$, the multifield bounce equation \eqref{eqn:bounceeqn} becomes
	\begin{align}\label{eqn:bounceeqn4}
		\pdv[2]{\phi_i}{\rho}+\frac{3}{\rho} \pdv{\phi_i}{\rho}=\pdv{V^{\{4\}}(\phi_i)}{\phi_i}.
	\end{align}
	Before we analyse this equation deeper, it is helpful to perform Fubini's substitution \cite{Fubini:1976jm,Lee:1985uv} of the fields $\bm\psi = \rho \bm\phi $ and radial coordinate $\rho=e^t$, where boldface denotes vector quantities. After these substitutions, the above equation reduces to a conservative form 
	\begin{align}
		\pdv[2]{\psi_i}{t} = \psi_i + \pdv{V^{\{4\}}(\psi_i)}{\psi_i}\,.
	\end{align}
	We can thus interpret it in terms of a classical system with Hamiltonian given by
	\begin{align}\label{eqn:bounceham}
		\mathcal{H} = \frac{1}{2}\bm\pi^2 - \frac{1}{2}\bm\psi^2 - V^{\{4\}}(\bm\psi)
	\end{align}
	and the initial condition ${\bm \psi}(-\infty)=\dot{\bm \psi}(-\infty) =0$. Here, $\bm\pi$ is the canonical conjugate momentum to the spatial variables $\bm\psi$. While the boundary condition $ \bm\phi(\infty)= \bm\phi_\text{FV}=0 $ is trivially satisfied for any finite value $\bm\psi(\infty)$, for this system to have a finite action, and thus a nonzero contribution to the tunneling rate, we further impose the boundary condition $\bm\psi(\infty)=0$. 
	
	If we now write the fields in polar form $\bm\psi=(\|\bm\psi\|,\bm\theta)$ it becomes clear that due to the homogeneity of the potential $V^{\{4\}}(\lambda\bm\psi)=\lambda^4 V^{\{4\}}(\bm\psi)$, the potential separates into a radial and angular component $V^{\{4\}} = \|\bm \psi\|^4 \mathcal{V}(\bm\theta)$. Since the only term in $\frac{1}{2}\bm\psi^2+V^{\{4\}}$ with angular dependence is $V^{\{4\}}$, and by the equations of motion $\dot{\bm\theta}\propto {\bm\pi}_{\bm\theta},\ \dot{\bm\pi}_{\bm\theta}\propto-\pdv{\mathcal{H}}{\bm\theta} $, any particle subject to $\mathcal{H}$ on the radial line $\left\{(\|\bm\psi\|,\theta_*): \|\bm\psi\| \in\mathbb{R}, \left.\frac{d\mathcal{V}}{d\bm\theta}\right|_{\bm\theta_*}=0\right\}$ will remain on the line, given that ${\bm\pi}_{\bm\theta}=0$ at the initial condition. For the time being, we assume the solution to sit on such a line and will discuss deviations from this subset later. 
    
	Since we have restricted $\bm\phi\propto\bm\psi$ to one dimension, the one-dimensional solution can then be immediately applied to give equation \eqref{eqn:bounceactioneff} where
	\begin{align}\label{eqn:lambdaeff}
		\lambda_\text{eff} = 4\mathcal{V}(\bm\theta_*)\comma \left.\frac{d\mathcal{V}}{d\bm\theta}\right|_{\bm\theta_*}=0.
	\end{align}
    Out of all the possible solutions $\bm\theta_*$, then, the one that gives the minimum $\lambda_\text{eff}$ provides the exponentially dominant tunneling rate. 
	
	Notably, the condition for this radial bounce to exist, is precisely the condition that the potential $V^{\{4\}}$ admits a lower vacuum. To see this, note that the boundary conditions below Eq.~\eqref{eqn:bounceham} mean that for a radial bounce to exist, the potential $- \frac{1}{2}\bm\psi^2 - V^{\{4\}}(\bm\psi)$ must have a valid turning point $\bm\psi_*$ such that $- \frac{1}{2}\bm\psi_*^2 - V^{\{4\}}(\bm\psi_*)=0$. Otherwise, the particle will continue to move outward and the boundary conditions at $\pm \infty$ cannot be simultaneously satisfied. Since $V^{\{4\}}(\bm\psi_*)$ is quartic, a turning point exists if and only if $\mathcal{V}(\bm\theta_*)<0$ for some $\bm\theta_*$. As such, the necessary requirements for a radial bounce to exist
	\begin{align}
	 	\left.\frac{d\mathcal{V}}{d\bm\theta}\right|_{\bm\theta_*}=0\comma \mathcal{V}(\bm\theta_*)<0
	\end{align}
	are exactly the requirements for a quartic potential to be unbounded from below. For renormalization scales $\mu \gg v$, this is exactly the condition for the effective potential to develop a lower vacuum. This suggests that the radial bounces are indeed the physically significant solutions to the vacuum decay rate. 
	
	Of course, other solutions to Eq. \eqref{eqn:bounceham} satisfying the boundary conditions do exist. However, we claim that if any nonradial solution $\gamma$ exists, then its contribution to the decay rate must be negligible. To see this, note that
	\begin{align}
		B &= \int d^4x \left[\frac{1}{2}(\dot{\bm\phi})^2+\frac{1}{2}(\nabla\bm\phi)^2+V^{\{4\}}(\bm\phi)\right] \nonumber\\
		&=2\pi^2 \int dt \left[\frac{1}{2}\dot{\bm\psi}^2+\frac{1}{2}\bm\psi^2+V^{\{4\}}(\bm\psi)\right]\nonumber \\
		&=2\pi^2 \int_\gamma ds \left[\sqrt{\bm\psi^2+2\bm\psi^4\mathcal{V}(\bm\theta)}\right]\,.
	\end{align}

	\begin{figure*}[t]
		\resizebox{1\textwidth}{!}{
			\includegraphics[height=0.3\textwidth]{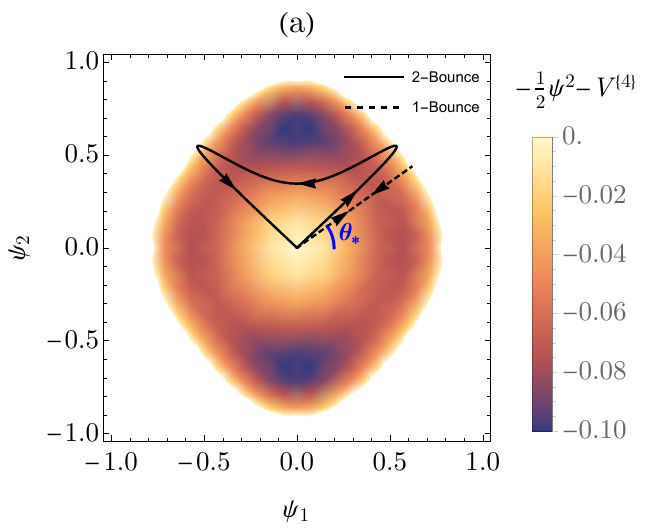}
			\includegraphics[height=0.3\textwidth]{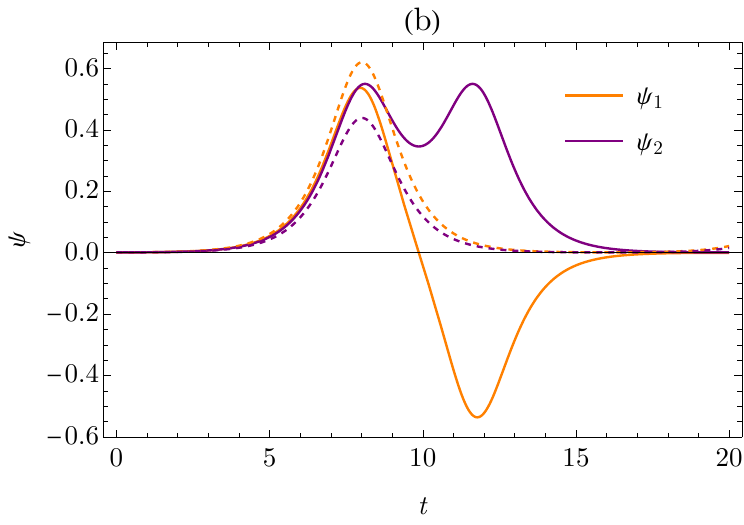}
		}
		\caption{Example bounce solutions for the two-scalar model of \eqref{eqn:twoscalar}, with $\lambda_1=-3.2,\lambda_2=-2.4,$ and $\lambda_3 = -4$. 
        (a) The trajectory of a radial bounce and a nonminimal bounce solution. The color of the heatmap is determined by the effective potential appearing in the Hamiltonian \eqref{eqn:bounceham}. (b) The $\psi_1$ and $\psi_2$ components of the two trajectories plotted against time. The dashed (solid) line denotes the 1-bounce (2-bounce) solution. 
        The radial 1-bounce solution not only minimizes the value of $\frac{1}{2}\psi^2+V^{\{4\}}$, but also takes the shortest distance to return to the origin, thereby minimizing the bounce action \eqref{eqn:bouncegeo}.}
		\label{fig:bounce}
	\end{figure*}
    Here, $ds = \|d\bm\psi\|$ is the line element. For all $\|\bm\psi\|$, the square root takes its minimum at the minimum of $\mathcal{V}(\bm\theta)$, which we denote $\bm\theta_*$. Therefore, the action of any nonradial bounce is necessarily bounded from below by the action of the same path evaluated at a fixed angle $\bm\theta_*$. This path is no longer guaranteed to be stationary. However, its action must still be bounded from below by a classical path $\gamma_*$ constrained along $\bm\theta_*$:
	\begin{align}\label{eqn:bouncegeo}
		\int_\gamma ds \left[\sqrt{\bm\psi^2+2\bm\psi^4\mathcal{V}(\bm\theta)}\right]&\geq \int_\gamma ds \left[\sqrt{\bm\psi^2+2\bm\psi^4\mathcal{V}(\bm\theta_*)}\right]\nonumber \\
		&\geq \int_{\gamma_*} ds \left[\sqrt{\bm\psi^2+2\bm\psi^4\mathcal{V}(\bm\theta_*)}\right]\,.
	\end{align}
	Since the decay rate contribution of each path is proportional to $e^{-B}$, we conclude that the radial bounce indeed provides the exponentially dominant contribution.
	
	Numerically, the situation is illustrated in Figure \ref{fig:bounce}, using a toy model with two scalars with a potential 
    \begin{align}\label{eqn:twoscalar}
        V = \frac{1}{4}\lambda_1 \phi_1^4 + \frac{1}{4}\lambda_2 \phi_2^4 + \frac{1}{2}\lambda_3 \phi_1^2 \phi_2^2.
    \end{align}
    Here, we plotted the lowest-action nonradial bounce we numerically found, along with a radial bounce. Already from inspection, it is clear that the radial bounce contributes a smaller bounce action compared to the nonradial bounce -- see the figure caption for more details.
	
\section{Tunneling Rates for Biquadratic Theories}
\label{Sec:tunnelingrates}
The previous section has dealt with the general theory for large-field vacuum decay for theories with multiple scalars. However, equation \eqref{eqn:lambdaeff} is still cumbersome to calculate for theories with more than 3 or 4 scalar fields. In this section, we present a straightforward analytical solution for when the quartic part of the potential can be written as a biquadratic form:
	\begin{align}
		V^{\{4\}} =\sum_{ij} \Lambda_{ij}\phi_i^2 \phi_j^2\,,
	\end{align}
    where $\Lambda_{ij}$ are the quartic couplings and $\phi_{i,j}$ are real scalar fields. Note that such a form is common in phenomenologically relevant models (see e.g. the 2HDM+a model below) and, as we will see later with the 3-3-1 model in Section~\ref{Sec:3-3-1}, the framework developed in this section is useful even for models that do not seem biquadratic at first glance.
    
	When the potential is biquadratic, minimising $\mathcal{V}$ is equivalent to minimising the quadratic form subject to constraints
	\begin{align}\label{eqn:constrainedproblem}
		\lambda_\text{eff} =4\min (\bm{u}^T \bm\Lambda \bm{u})\comma u_i>0\comma \sum_i u_i=1\,,
	\end{align}
    where the case of vanishing $u_i$ will be discussed later.
	The condition $\sum_i u_i =1$, which corresponds to pulling out the factor $\|\bm \psi \|^ 4$ of the potential to arrive at ${\cal V}$, can be implemented using a Lagrange multiplier $\lambda$ with the Lagrangian given by
	\begin{align}\label{eqn:lagrange}
		L = \bm{u}^T\bm\Lambda \bm{u}- \lambda(\bm{u}^T \bm{j} - 1)\,.
	\end{align}
	Here, we introduced a vector with unit entries $j_i = 1$.
    The remaining condition $\forall i,u_i>0$ can then be checked manually for each solution, discarding those that violate the constraint. The condition $\pdv{L}{u} = 0$ then directly gives
	\begin{align}
		2\bm\Lambda \bm{u}-\lambda\bm{j}=0 \implies \bm{u}=\frac{1}{2}\lambda \bm\Lambda^{-1}\bm{j}\,.
	\end{align}
	Taking the inner product on both sides with $\bm{j}$ and using the fact that $\bm{u}\cdot\bm{j}=1$,
	\begin{align}
		\lambda = \frac{2}{\bm{j}^T\bm\Lambda^{-1}\bm{j}} \implies  \bm{u}=\frac{\bm\Lambda^{-1}\bm{j}}{\bm{j}^T\bm\Lambda^{-1}\bm{j}}.
	\end{align}
	Finally, we find
	\begin{align}\label{eqn:lambdaeffbq}
		\lambda_\text{eff} = 4 \bm{u}^T \bm{\Lambda} \bm{u}= \frac{4}{\bm{j}^T\bm\Lambda^{-1}\bm{j}}.
	\end{align}
	So far, we have only discussed solutions $\bm{u}$ in the interior of the simplex $u_i>0, \sum_i u_i = 1$. However, solutions on the boundaries, where $u_{k'} = 0$ for all $k'\in K' \subset\{1,\cdots,n\}$ may also exist. In such cases, instead of adding another constraint to \eqref{eqn:constrainedproblem}, note that when $u_{k'}$ is fixed to zero, the $k'$ entry no longer contributes to the Lagrangian. Let us define $K = \{1,\cdots,n\}\setminus K'$. Then the submatrix $\bm\Lambda_K$, with all $K'$ rows and columns removed from $\bm\Lambda$, can be used in place of $\bm\Lambda$ to solve the problem \eqref{eqn:constrainedproblem} constrained on the boundary. The set of all radial bounces is then given by \eqref{eqn:bounceactioneff} with
	\begin{align}\label{eqn:completelambdaeff}
		\left\{ \lambda_\text{eff}^K = \frac{4}{\bm{j}^T\bm\Lambda^{-1}_K\bm{j}} : K \subset \{1,\cdots,n\} \comma \bm{u}_K \propto\bm\Lambda^{-1}_K\bm{j} >0\right\}
	\end{align} 
	and the leading contribution to the tunneling rate is given by the minimum of the above set.
	
	There is a subtlety that appears when $\bm\Lambda_K$ is not invertible. In such cases, there are two scenarios: $\bm{j}$ is not in the image of $\bm{\Lambda}_K$ in which no solution needs to be considered, or $\bm{j}$ is indeed in the image. In the latter scenario, since $\bm\Lambda_K$ is symmetric, the nullspace $\mathcal{N}(\bm\Lambda_K)$ of $\bm{\Lambda}_K$ is orthogonal to $\bm{j}$ in the image. As such, for any particular solution $\bm{v}_p$ with $\Lambda \bm{v}_p = \bm{j}$, the complete solution $\bm{v}\in\bm{v}_p + \mathcal{N}(\bm\Lambda_K)$ also satisfies the constraint $\bm{u}\cdot\bm{j}=\frac{1}{2}\lambda\bm{v}\cdot\bm{j}=1$. In addition, all solutions $\bm{v}$ give the same bounce action $\lambda_\text{eff}^K = \frac{4}{\bm{j}\cdot\bm{v}}$, so it suffices to consider any one solution. Then, since $\mathcal{N}(\bm\Lambda_K)$ is not trivial, there must exist a solution $\bm{u}$ on the boundary of the simplex. So we can find a subset $\tilde{K}\subset K$ for which $\Lambda_{\tilde{K}}$ is invertible, and whose solution $\tilde{\bm{u}}$ is in the complete solution of the original problem $\tilde{\bm{u}} \in \bm{u}_p + \mathcal{N}(\bm\Lambda_K)$. Therefore, the set \eqref{eqn:completelambdaeff} captures all the solutions we need.
	
	In practice, the inversion operation makes the analytical expression in terms of the components $\Lambda_{ij}$ too complicated when $\dim\bm\Lambda\gtrsim 6$. Evaluating~\eqref{eqn:lambdaeffbq} in Mathematica for $1\leq \dim\bm\Lambda \leq 8$ shows that the number of arithmetic operations $N_{\text{op}}$ in the unsimplified expression scales as $N_{\text{op}} \sim 0.16 \cdot 7.9^{\dim\bm\Lambda}$, which quickly becomes intractable. However, $\bm{\Lambda}^{-1}\bm{j}$ can be very efficiently calculated numerically using standard algorithms for solving symmetric linear systems, for hundreds of scalar fields if needed. The main remaining bottleneck is thus enumerating all the possible subsets $K$ of fields to consider in \eqref{eqn:completelambdaeff}, whose optimization will depend on the specifics of the model.

	\bgroup
	\def\arraystretch{1.5}
	\begin{table*}[!ht]
		\centering
		\begin{ruledtabular}
			\begin{tabular}{c|c}
				\hspace*{1cm}$\dim \bm\Lambda$ \hspace*{1cm} & $\frac{1}{4}\lambda_\text{eff}$  \hspace*{1cm} \\ 
				\hline
				1& $\Lambda _{1 1}$  \hspace*{1cm}\\
				2& $\frac{\left(\Lambda _{1 1} \Lambda _{2 2}-\Lambda _{1 2}^2\right)}{\Lambda _{1 1}-2 \Lambda _{1 2}+\Lambda _{2 2}}$ \hspace*{1cm}\\
				3& $\frac{ \left(\Lambda _{3 3} \Lambda _{1 2}^2-2 \Lambda _{1 3} \Lambda _{2 3} \Lambda _{1 2}+\Lambda _{1 3}^2 \Lambda _{2 2}+\Lambda _{1 1} \left(\Lambda _{2 3}^2-\Lambda _{2 2} \Lambda _{3 3}\right)\right)}{\Lambda _{1 2}^2-2 \left(\Lambda _{1 3}+\Lambda _{2 3}-\Lambda _{3 3}\right) \Lambda _{1 2}+\Lambda _{1 3}^2+\Lambda _{2 3}^2+2 \Lambda _{1 3} \left(\Lambda _{2 2}-\Lambda _{2 3}\right)-\Lambda _{2 2} \Lambda _{3 3}-\Lambda _{1 1} \left(\Lambda _{2 2}-2 \Lambda _{2 3}+\Lambda _{3 3}\right)}$ \hspace*{1cm} \\
			\end{tabular}
		\end{ruledtabular}
		\caption{General expressions for $\lambda_\text{eff}$ for $\dim \bm\Lambda = 1,2,3$. }
		\label{tab:lambdaeff}
	\end{table*}
	\egroup

	For reference, we provide the general expressions for the first three $\lambda_\text{eff}$ in Table \ref{tab:lambdaeff}. As we will see, this is actually all we need to fully describe the bounces in the 2HDM+a and 3-3-1 examples.

\section{2HDM+a Example}
\label{Sec:2HDMa}

\subsection{Model and Conventions}
\label{Sec:2hdmamodel}
The Two-Higgs-Doublet-Plus-Pseudoscalar (2HDM+a) model was introduced in \cite{Bauer:2017ota} as a simplified model for dark matter. We will use this model as an example use case for our main result \eqref{eqn:completelambdaeff}, and derive some preliminary results on the effect of vacuum decay on the phenomenology of the model. The scalar potential is given by
	\begin{align}\label{eqn:2hdmapotentials}
		V_\text{2HDM} &= m_{11}^2 \Phi_1^\dagger \Phi_1 - (m_{12}^2 \Phi_1^\dagger \Phi_2 +\text{h.c.})+ m_{22}^2 \Phi_2^\dagger \Phi_2 \nonumber\\
		&+\frac{1}{2}\lambda_1 (\Phi_1^\dagger \Phi_1)^2+\frac{1}{2}\lambda_2 (\Phi_2^\dagger \Phi_2)^2 +\lambda_3 (\Phi_1^\dagger \Phi_1)(\Phi_2^\dagger \Phi_2) \nonumber\\
		&+\lambda_4 (\Phi_1^\dagger \Phi_2)(\Phi_2^\dagger \Phi_1)+ \frac{1}{2}(\lambda_5(\Phi_1^\dagger \Phi_2)^2 + \text{h.c.})\\
		V_\text{2HDM+a} &= V_\text{2HDM} + \frac{1}{2} m_{a_0}^2 a_0^2 + \frac{\lambda_a}{4}a_0^4 \nonumber \\
		&+ ( \iu \mu_{12a} \Phi_1^\dagger \Phi_2a_0 +\text{h.c.}) \nonumber\\ \label{eqn:2hdmapotential}
		&+\frac{1}{2}\lambda_{11a} \Phi_1^\dagger \Phi_1 a_0^2
		+\frac{1}{2}\lambda_{22a} \Phi_2^\dagger \Phi_2 a_0^2.
	\end{align}
    Here, we have imposed a softly-broken $\mathbb{Z}_2$ symmetry $\Phi_2\mapsto-\Phi_2$, to naturally suppress FCNCs \cite{Branco:2011iw} and an explicit CP symmetry, keeping $m_{12}^2, \lambda_5$ and $\mu_{12a}$ real.
	The pseudoscalar $a_0$ is also coupled to a fermionic dark matter candidate $\chi$ via a yukawa interaction
	\begin{align}
		\mathcal{L}_\text{2HDM+a} \supset -i y_\chi \bar\chi \gamma^5 \chi a_0.
	\end{align}
	After Electroweak Symmetry Breaking, this scalar sector contains two neutral scalars $h,H$, a charged scalar $H^\pm$, and two pseudoscalars $a,A$. We adopt the convention for the pseudoscalar mixing angle as
	\begin{align}
		\begin{pmatrix}
			A_0\\a_0
		\end{pmatrix} = \begin{pmatrix}
            \cos\theta & -\sin\theta\\ \sin\theta & \cos\theta
        \end{pmatrix} \begin{pmatrix}
		A\\a
		\end{pmatrix}\,,
	\end{align}
    where the fields with $\square_0$ subscripts are in the interaction basis as written in \eqref{eqn:2hdmapotential}, and the fields without subscript are the mass eigenstates. 
	This results in a change of variables given by
	\begin{align}\label{eqn:2hdma-change}
		\mu_{12a} &= \frac{(m_a^2-m_A^2)\sin(2\theta)}{2v},\nonumber\\ m_{A_0}^2 &= m_A^2 \cos^2(\theta)+m_a^2\sin^2(\theta).
	\end{align}
    Here, $m_{A_0}^2 := \left.\pdv[2]{V_\text{2HDM}}{A_0}\right|_v
    $ is defined for convenience\footnote{In the 2HDM+a, the $m_A^2$ appearing in the 2HDM expressions in \cite{Branco:2011iw,Ferreira:2019bij} must be replaced with $m_{A_0}^2$.}.  
	The remaining 2HDM parameters are also replaced using the expressions provided in \cite{Branco:2011iw, Ferreira:2019bij}. This leads to the following set of input parameters: 
	\begin{align}
		m_h, v, m_H, m_{H^\pm}, m_A, \beta, \alpha-\beta,\lambda_1, m_a, \theta, \lambda_{11a}, \lambda_{22a}, \lambda_a.
	\end{align}
	Here, $\alpha,\beta$ are the conventional mixing angles in the 2HDM.

\subsection{Vacuum Metastability}
\label{Sec:2hdmametastability}
	Exploiting the $\mathrm{U}(2)$ symmetry, the 2HDM fields can be written in the standard form $\Phi_1 = \frac{1}{\sqrt{2}} \begin{pmatrix}\phi_1\\ \phi_2+i\phi_3\end{pmatrix},\Phi_2 =  \frac{1}{\sqrt{2}}\begin{pmatrix}0\\ \phi_4\end{pmatrix}$ \cite{Branco:2011iw}. The potential thus reduces to being dependent on five fields $(\phi_1,\cdots,\phi_4,a)$. The quartic part of the potential can then be written as a biquadratic form with
	\begin{align}\label{eqn:2hdmalambda}
		\bm\Lambda=\frac{1}{8}\left(
		\begin{array}{ccccc}
			\lambda_1 & \lambda_1 & \lambda_1 & \lambda_3 & \lambda_{11a} \\
			\lambda_1 & \lambda_1 & \lambda_1 & \lambda _{345}^+ & \lambda_{11a} \\
			\lambda_1 & \lambda_1 & \lambda_1 & \lambda _{345}^- & \lambda_{11a} \\
			\lambda_3 & \lambda _{345}^+ &\lambda _{345}^- & \lambda_2 & \lambda_{22a} \\
			\lambda_{11a} & \lambda_{11a} & \lambda_{11a} & \lambda_{22a} & 2 \lambda_a \\
		\end{array}
		\right).
	\end{align}
	Here, $\lambda_{345}^\pm = \lambda_3+\lambda_4 \pm \lambda_5,$ and the fields are ordered as $\bm{u} = (\phi_1^2,\cdots,\phi_4^2,a_0^2)$.
	
	Luckily, for any $\bm\Lambda_K$ with $\|K\|>1$, if there exists a set of rows $R\subset K$ and a set of columns $C\subset R$ such that $\|C\|+\|R\|> \|K\|$ and the columns $C$ of $\bm\Lambda$ restricted to rows $R$ are all proportional to $\bm{j}$, then either $\det\bm\Lambda_K = 0$ or the unique solution has $u_i=0$ for some $i$. Thus, the submatrix can be discarded, as discussed in Sec.~\ref{Sec:tunnelingrates}. 

   \emph{Proof.} Suppose $\bm\Lambda_K$ is such that $\det\bm\Lambda_K \neq 0$, so that the columns are independent, and $\|C\|+\|R\|>\|K\|$ as described above. Now consider the linear equation $\bm{\Lambda}_K\bm{v}=\bm{j}$, with $v_i=0$ for all $i\not\in C$. Since the rows $R$ restricted to C are all identical, we are effectively solving a system of $\|K\|-\|R\|+1$ equations with $\|C\|$ unknowns. Finally, since the columns $C$ are linearly independent, and there are at least as many unknowns as equations $\|C\|\geq \|K\|-\|R\|+1$, we can find a solution to this system. Note that this solution automatically has $v_i =0 $ for all $i\not\in C$, and since $\det\bm\Lambda_K \neq 0$, we cannot have $C=K.${\null\nobreak\hfill\ensuremath{\square}}

    This is the case for all submatrices $\bm\Lambda_K$ with $\|K\|> 3$.
	As such, all the bounces for the 2HDM+a can be found by enumerating through all submatrices with $\|K\|\leq 3$, and using the general expressions in Table \ref{tab:lambdaeff}.  
    Of these submatrices, 15 of them satisfy the requirement $\|C\|+\|R\|\leq\|K\|$. However, some of these are duplicates so in the end we have only 11 unique values for $\lambda_\text{eff}$ to care about. The 11 unique submatrices are provided here for completeness:

    \begin{align} \label{eqn:2hdmamatrices}
        &\left(
\begin{array}{c}
 \lambda_1 \\
\end{array}
\right),\left(
\begin{array}{c}
 \lambda_2 \\
\end{array}
\right),\left(
\begin{array}{c}
 2 \lambda_{a} \\
\end{array}
\right),\left(
\begin{array}{cc}
 \lambda_1 & \lambda_3 \\
 \lambda_3 & \lambda_2 \\
\end{array}
\right),\left(
\begin{array}{cc}
 \lambda_1 & \lambda_{11a} \\
 \lambda_{11a} & 2 \lambda_{a} \\
\end{array}
\right),
 \nonumber \\ &\left. 
\left(
\begin{array}{cc}
 \lambda_1 & \lambda_{345}^+ \\
 \lambda_{345}^+ & \lambda_2 \\
\end{array}
\right),\left(
\begin{array}{cc}
 \lambda_1 & \lambda_{345}^- \\
 \lambda_{345}^- & \lambda_2 \\
\end{array}
\right),\left(
\begin{array}{cc}
 \lambda_2 & \lambda_{22a} \\
 \lambda_{22a} & 2 \lambda_{a} \\
\end{array}
\right),
\right. \nonumber \\ &\left. 
\left(
\begin{array}{ccc}
 \lambda_1 & \lambda_3 & \lambda_{11a} \\
 \lambda_3 & \lambda_2 & \lambda_{22a} \\
 \lambda_{11a} & \lambda_{22a} & 2 \lambda_{a} \\
\end{array}
\right),\left(
\begin{array}{ccc}
 \lambda_1 & \lambda_{345}^+ & \lambda_{11a} \\
 \lambda_{345}^+ & \lambda_2 & \lambda_{22a} \\
 \lambda_{11a} & \lambda_{22a} & 2 \lambda_{a} \\
\end{array}
\right),
\right. \nonumber \\ &
\left(
\begin{array}{ccc}
 \lambda_1 & \lambda_{345}^- & \lambda_{11a} \\
 \lambda_{345}^- & \lambda_2 & \lambda_{22a} \\
 \lambda_{11a} & \lambda_{22a} & 2 \lambda_{a} \\
\end{array}
\right)\,,
    \end{align}
	 from which the $\lambda_\text{eff}$ can be obtained straightforwardly. 
     
\subsection{Theoretical Constraints} 
\label{Sec:theconstr}
We consider three types of theoretical constraints: perturbativity (P), perturbative unitarity (PU), and boundedness from below (BFB). For perturbativity, we simply require $\forall i:\lambda_i<4\pi$. The inequalities for PU are obtained by requiring that the eigenvalues of the $2\to 2$ scattering matrix be smaller than the unitarity limit \cite{LHCDarkMatterWorkingGroup:2018ufk}
	\begin{align}
		|\lambda_{11a,22a}|&< 4\pi, \nonumber \\|\lambda_3\pm\lambda_{4,5}|&<8\pi,  \nonumber\\
		\lambda_1+\lambda_2\pm \sqrt{(\lambda_1-\lambda_2)^2+4\lambda_{4,5}^2}&<16\pi,\nonumber\\
		|x_i|&<8\pi,
	\end{align}
	where $x_i$ are the roots of the following polynomial:
	\begin{align}
		&x^3 - 3(\lambda_a +\lambda_1+\lambda_2)x^2 +  (9\lambda_1\lambda_a +9\lambda_2\lambda_a\nonumber\\&-\lambda_{11a}^2-\lambda_{22a}^2-4\lambda_3^2-4\lambda_3\lambda_4-\lambda_4^2+9\lambda_1\lambda_2)x\nonumber\\&+ 3\lambda_{22a}^2\lambda_1+ 3\lambda_{11a}^2\lambda_2-\lambda_{11a}\lambda_{22a}(4\lambda_3-2\lambda_4)\nonumber\\&+(-27\lambda_1\lambda_2+12\lambda_3^2+12\lambda_3\lambda_4+3\lambda_4^2)\lambda_a.
	\end{align}
	Finally, to keep the tree-level potential bounded from below, we will simply require the implication $\bm{u}_K>0 \implies \lambda_\text{eff}^K>0$ at tree-level for all subsets $K$, c.f. \eqref{eqn:lambdaeffbq}. This is equivalent to the copositivity criterion in \cite{Kannike:2012pe}.

\subsection{Experimental Constraints} 
\label{Sec:2hdmaexpconstr}
Given that the primary focus of this work is to demonstrate the application of the bounce action formalism for tunneling vacua, rather than to perform a comprehensive phenomenological analysis of the 2HDM+$a$ model, we will restrict ourselves to a minimal set of experimental constraints. In particular, we consider electroweak precision observables, namely the $S$, $T$, and $U$ parameters \cite{Kanemura:2011sj,Haber:2010bw}, since these depend to leading order only on the scalar particle masses and mixings and therefore provide a robust test of the extended scalar sector, performing an analysis along the lines of \cite{Arcadi:2023smv}. We will not specify a particular type of 2HDM, as the couplings to light fermions are not relevant for our analysis\footnote{The Yukawa type typically changes very weakly the RG running of $\lambda_\text{eff}$.}. Consequently, we do not include flavour physics constraints. Regarding Higgs measurements, we make use of the results of Ref.~\cite{ATLAS:2024lyh}, adopting their Type I scenario, since it allows for a larger viable region of parameter space compared to the other Yukawa structures. Finally, we do not include collider bounds, as the 2HDM+$a$ is typically only weakly constrained by direct searches.

\subsection{Results}
\label{Sec:2hdmaresults}

\begin{figure*}[t]
		\resizebox{1\textwidth}{!}{
			\includegraphics[height=0.3\textwidth]{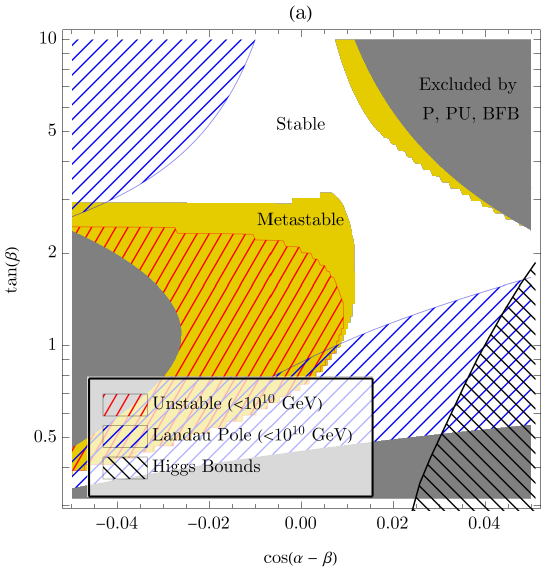}
			\includegraphics[height=0.3\textwidth]{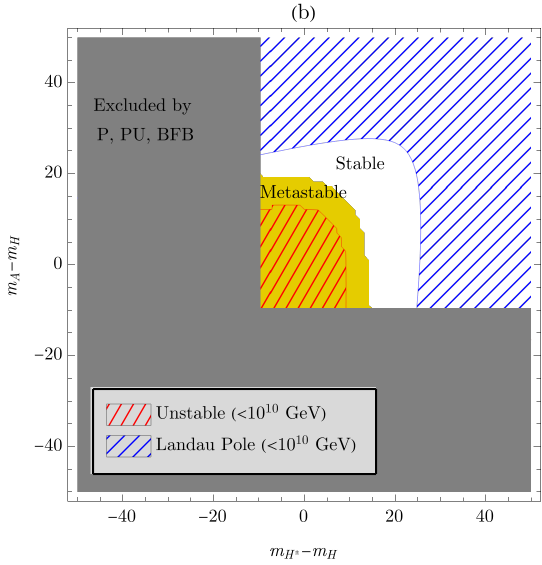}
		}
		\caption{Parameter scans of the 2HDM+a model phase diagram (a) in the $\cos(\alpha-\beta)\times \tan(\beta)$ plane and (b) in the $ m_{H^\pm}-m_H\times m_A-m_H $ plane. All other parameters are fixed as given in \eqref{eqn:2hdm-paramchoices}. Gray regions are excluded by tree level perturbativity, unitarity, and bounded-from-below constraints. All high-scale validity constraints are only considered up to a cutoff scale $10^{10}\,$GeV. Constraints due to instabilities beyond this scale are ignored. See text for more details.}
		\label{fig:scan-2hdma-cab-tb}
\end{figure*}

Since we are mainly interested in the validity of the 2HDM+a model up to energy scales of $\Lambda \gg v$, to be consistent, we must also consider constraints due to a low-energy Landau pole. Similar to the vacuum decay constraint, the presence of a Landau pole at a scale $\Lambda$ means that the model is no longer a good description of physics at energies $\gtrsim
\Lambda$. However, the implications of Landau pole constraints are slightly weaker than vacuum decay constraints: While the presence of a lower energy vacuum at $\Lambda$ implies that new physics must kick in at scales $\ll \Lambda$ to stabilize the vacuum, a Landau pole only tells us that perturbation theory is no longer valid (for example, due to condensation). As a result, one should take the Landau pole constraints less seriously than the vacuum decay constraints.

To compare with experimental constraints, we scanned both the $\cos(\alpha-\beta)$ versus $\tan\beta$ and $m_{H^\pm}-m_H$ versus $m_A-m_H $ planes. The results are shown in Fig.~\ref{fig:scan-2hdma-cab-tb}. For the parameters kept constant, we have chosen to adopt the following choices:
\begin{align}\label{eqn:2hdm-paramchoices}
    m_H = m_{H^\pm}=m_A = 600\,\mathrm{GeV}, \nonumber\\ \cos(\alpha-\beta)=0\comma  \tan\beta=2 \comma
    \lambda_1 = 0.6,  \nonumber\\ m_a=150\,\text{GeV}\comma \theta=0\nonumber\\
    \lambda_a = \lambda_{11a}=\lambda_{22a}=0 \comma y_\chi =0\,.
\end{align}
Note that in both scans here, the vacuum decay constraints are mostly dominated by 2HDM contributions $\lambda_\text{eff} = \lambda_2$. This is caused by the large contribution to the running due to the sizable yukawa coupling between the top quark and $\Phi_2$, and this contribution is relatively insensitive to the light pseudoscalar sector~\eqref{eqn:2hdmapotential}\footnote{More precisely, it is sensitive to $m_{A_0}^2$, so if we parameterize the model in terms of $m_A,m_a,\sin\theta$, then it is indirectly sensitive to $m_a,\sin\theta$. However, to leading order, this only amounts to a uniform shift in the allowed region of $m_A$. }. Similarly, vacuum decay contributions from this sector \eqref{eqn:2hdmapotential} are relatively insensitive to the bare 2HDM parameters. We thus analyse the two sectors separately: we focus on the 2HDM sector first and consider the light pseudoscalar sector \eqref{eqn:2hdmapotential} later. As such, we set $\theta = \lambda_a = \lambda_{11a}=\lambda_{22a}=0$ and $y_\chi=0$ for convenience. The constraints presented in Fig.\,\ref{fig:scan-2hdma-cab-tb} thus also apply directly to 2HDM models. Note that the pure 2HDM case has been previously studied in \cite{Chakrabarty:2016smc} by analyzing the $\lambda_2$ tunneling direction. Here, we extend their analysis by explicitly checking all the conditions in \eqref{eqn:2hdmamatrices} and Table.~\ref{tab:lambdaeff}. Indeed, we find that other tunneling directions do not contribute to the constraints unless we tune $\tan\beta \gg 10$ in a non-Type-I 2HDM. Finally, the rationale behind $\lambda_1=0.6$ in our benchmark is that smaller values of $\lambda_1$ tend to make vacuum decay constraints stricter, while a too-large value of $\lambda_1$ would lead to Landau pole instabilities. We thus choose $\lambda_1 = 0.6$ to show the typical behavior of both vacuum decay and Landau pole constraints. 

From Fig.~\sref{fig:scan-2hdma-cab-tb}{(a)}, we see that Higgs bounds are only relevant for small $\tan\beta$, and are in any case mostly overshadowed by the Landau pole constraints. Similarly, concerning the mass splittings analyzed in Fig.~\sref{fig:scan-2hdma-cab-tb}{(b)}, current bounds on the electroweak precision observables constrain $m_H-m_{H^\pm}$ and $m_H-m_A$ only down to $\lesssim 100\,$GeV, being much weaker than the Landau pole bounds. In summary, all current experimental bounds considered here barely probe the parameter space of high-scale-valid 2HDMs.

Another notable feature in Fig.~\sref{fig:scan-2hdma-cab-tb}{(b)} is that small mass differences of $\sim 10\,$GeV have the potential to stabilize the vacuum. Indeed, the parameters in Fig.~\sref{fig:scan-2hdma-cab-tb}{(a)} excluded by vacuum decay can still be allowed for non-vanishing mass differences. Given the opposite behavior of the vacuum decay and electroweak constraints, we further expect that more precise measurements of the STU parameters may work in concert with our vacuum decay constraints to rule out considerable parts of the low $\tan\beta$ region for all yukawa-types of 2HDMs. Note however that for the 2HDM+a, a nonzero $\sin\theta$ shifts this picture, as the most relevant parameter controlling perturbativity, vacuum decay, and Landau pole constraints is $m_{A_0}-m_H$. Therefore, larger values of $m_A-m_H \gtrsim 20\,$GeV are allowed at larger values of $\sin\theta$. 

\begin{figure*}[t]
		\resizebox{1\textwidth}{!}{
			\includegraphics[height=0.3\textwidth]{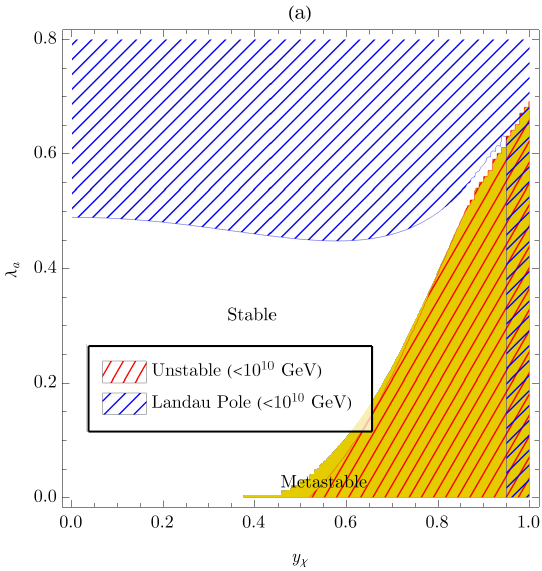}
			\includegraphics[height=0.3\textwidth]{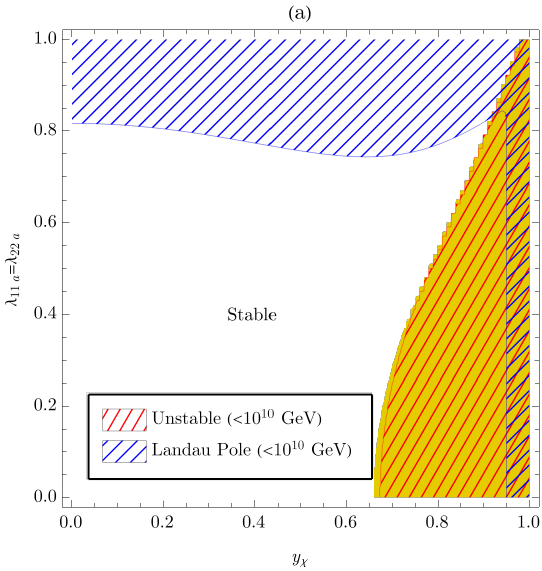}
		}
		\caption{Parameter scans of the 2HDM+a model phase diagram (a) in the $y_\chi \times \lambda_a$ plane and (b) in the $ y_\chi\times \lambda_{11a}=\lambda_{22a} $ plane. All other parameters are fixed as given in \eqref{eqn:2hdma-paramchoices}. All high-scale validity constraints are only considered up to a cutoff scale $10^{10}\,$GeV. Constraints due to instabilities beyond this scale are ignored.}
		\label{fig:scan-2hdma-yx-l}
\end{figure*}

We now turn to vacuum decay constraints independent from the $\lambda_\text{eff}=\lambda_2$ case considered above. To do this, we choose a point in Fig.\,\ref{fig:scan-2hdma-cab-tb} that is stable, and study the vacuum decay constraints we get by varying other parameters. For the parameter region introduced by the new light pseudoscalar $a$, we find that vacuum decay constraints are relatively uninteresting except for the parameter pairs ($\lambda_a$,$y_\chi$) and ($\lambda_{iia},y_\chi$). Both scans are shown in Fig.\,\ref{fig:scan-2hdma-yx-l}. For parameters kept constant, we have chosen to adopt the following:
\begin{align}\label{eqn:2hdma-paramchoices}
    m_H = m_{H^\pm}= 600\,\mathrm{GeV}\comma m_A = 690\,\mathrm{GeV}, \nonumber\\   \cos(\alpha-\beta)=0\comma \tan\beta=5 \comma
    \lambda_1 = 0.6, \nonumber\\  m_a=150\,\text{GeV}\comma \sin\theta=\frac{1}{2}\comma \lambda_a=\lambda_{11a}=\lambda_{22a}=0.2.
\end{align}
Note that unlike for Fig.\,\ref{fig:scan-2hdma-cab-tb}, we have chosen $\tan\beta = 5$ so that the 2HDM sector is stable, and we have taken $\lambda_{a} = \lambda_{11a}=\lambda_{22a}=0.2$ small and positive, to make sure the vacuum decay constraints are not too strict, and to avoid spurious results caused by floating point errors. For the scan in Fig.~\sref{fig:scan-2hdma-yx-l}{(b)}, we still assume $\lambda_{11a}=\lambda_{22a}$ to keep the process $\Gamma(H\to aa)$ small \cite{LHCDarkMatterWorkingGroup:2018ufk}. We have also chosen to set $\sin\theta=\frac{1}{2}$, to make more immediate contact with DM phenomenology \cite{LHCDarkMatterWorkingGroup:2018ufk}. As a result, to keep $m_{A_0} \approx m_H=m_{H^\pm}$, so that the 2HDM parameters are perturbative, we must set $m_A \approx 690\,$GeV. Note that both $\sin\theta$ and $m_A$, within their values allowed by PU, BFB, and Landau pole constraints, barely impact the vacuum stability constraints in Fig.\,\ref{fig:scan-2hdma-yx-l}. So the bounds here should apply approximately for any value of $\sin\theta$, as long as we keep $m_{A_0}\approx m_H$. %

Note that in both cases shown in Fig.\,\ref{fig:scan-2hdma-yx-l}, a larger value of the quartic couplings $\lambda_a,\lambda_{11a},\lambda_{22a}$ provides a more stable vacuum, while a higher value of the yukawa coupling $y_\chi$ destabilizes the vacuum. However, we cannot have $\lambda_a,\lambda_{11a},\lambda_{22a} \gtrsim 0.8$ without introducing a Landau pole before $10^{10}\,$GeV. This effectively limits the dark matter yukawa coupling $y_\chi$ to $\lesssim 0.8$ within the 2HDM+a model.

\section{3-3-1 Example}
\label{Sec:3-3-1}

\subsection{Model and Conventions}

In 3-3-1 models~\cite{Pisano:1992bxx,Frampton:1992wt,Singer:1980sw,Valle:1983dk} the SM gauge group is extended to $SU(3)_c \times SU(3)_L \times U(1)_X$, which leads to a set of interesting consequences. For example, these models (i) can explain the number of fermion generations via anomaly cancellation \cite{Frampton:1992wt,Pisano:1992bxx,Valle:1983dk} and naturally treat one quark generation differently, (ii) feature generically viable dark matter candidates \cite{Filippi:2005mt,deSPires:2007wat,Mizukoshi:2010ky,Ruiz-Alvarez:2012nvg,Kelso:2013nwa,Dong:2014esa,Dong:2015rka,Ferreira:2016uao,Arcadi:2017xbo,RodriguesdaSilva:2014gbi}, (iii) can implement light neutrino masses in a very natural way~\cite{Montero:2001ts,Chang:2006aa,Dong:2008sw}, as well as (iv) provide interesting ingredients to explain the matter-antimatter asymmetry of the universe~\cite{Phong:2013cfa,Phong:2014ofa,Huong:2015dwa,Borges:2016nne,Huang:2017laj,Dong:2017ayu,Boubakir:2021yho}.

The enlarged gauge symmetry can be spontaneously broken to the SM gauge group and finally to $U(1)_Q$ of electromagnetism (EM) in a two step process, along with generating viable fermion masses, via a scalar sector consisting of three $SU(3)_L$ triplets (even though other variants exist~\cite{Diaz:2003dk,Long:2024gyy})
\begin{equation}
  \rho = \begin{pmatrix}
       \rho_1 \\
       \rho_2 \\
       \rho_3
   \end{pmatrix},\
  \eta = \begin{pmatrix}
       \eta_1 \\
       \eta_2 \\
       \eta_3
   \end{pmatrix},\  
 \chi =  \begin{pmatrix}
       \chi_1 \\
       \chi_2 \\
       \chi_3
   \end{pmatrix}\,,
\end{equation}
with $U(1)_X$ charges $X_{\rho,\eta,\chi}$.
The first breaking $SU(3)_L \times U(1)_X \to SU(2)_L \times U(1)_Y$ is achieved for (see, e.g., \cite{Diaz:2003dk})
\begin{equation}
\label{eq:vchi}
\langle \chi \rangle = \begin{pmatrix}
       0 \\
       0 \\
       v_\chi
   \end{pmatrix}\,,
\end{equation}
with $Y=\beta \, \lambda_8/2 + X\, \mathbbm{1}$, where the famous $\beta$ parameter, defining different 3-3-1 variants, is given by $\beta = \sqrt 3 X_\chi$, while $\lambda_8$ is the eighth Gell-Mann matrix. The non-observation of the new gauge bosons of the broken symmetries suggests $v_\chi \gtrsim\,{\cal O}({\rm TeV})$.
The second breaking step $SU(2)_L \times U(1)_Y \to U(1)_Q$ can be achieved via the remaining $SU(3)_L$ triplets with $ X = \pm 1/2 - \beta/(2\sqrt3)$, embedding $SU(2)_L$ Higgs doublets with $Y=\pm1/2$ in their upper components, which feature vevs. Explicitly, we will chose
$X_\rho=1/2 - \beta/(2\sqrt3)$ and $X_\eta=-1/2 - \beta/(2\sqrt3)$, with EM-neutral $SU(2)_L$--breaking vevs
\begin{equation}
\label{eq:ver}
\langle \rho \rangle = \begin{pmatrix}
        0 \\
        v_\rho \\
        0 
   \end{pmatrix}\,,\
   \langle \eta \rangle = \begin{pmatrix}
        v_\eta \\
        0 \\
        0
   \end{pmatrix}\,.
\end{equation}
The symmetry breaking can be summarized as
\begin{equation}
\begin{split}
    SU(3)_L & \!\times\! U(1)_X  \xrightarrow{\langle \chi\rangle}SU(2)_L \!\times\! U(1)_Y \xrightarrow{\!\langle \rho\rangle, \langle \eta\rangle\!} U(1)_Q\,, \\
    Q&=\frac 1 2 \left(\lambda_3 + \beta \lambda_8 \right) + X \mathbbm{1}\,,\quad \beta=\sqrt3 X_\chi\,.
    \end{split}
\end{equation}

In the following, we will focus on a variant of this setup featuring $\beta = -1/ \sqrt{3}$, which naturally incorporates right-handed neutrinos (RHN) as the third components of the $SU(3)_L$ triplet leptons, known as the 3-3-1RHN model~\cite{Montero:1992jk,Foot:1994ym}.
The corresponding scalar potential reads
\cite{Diaz:2003dk,Montero:1992jk}
\begin{align}\label{eqn:331potential}
    V_{331} &= m_1^2\rho^\dagger \rho + m_2^2 \eta^\dagger \eta + m_3^2 \chi^\dagger \chi \nonumber\\
    &+ \lambda_1 (\rho^\dagger \rho)^2+ \lambda_2 (\eta^\dagger \eta)^2+ \lambda_3 (\chi^\dagger \chi)^2  \nonumber\\
    &+ \lambda_{12} (\rho^\dagger \rho)(\eta^\dagger \eta)+ \lambda_{13} (\rho^\dagger \rho)(\chi^\dagger \chi)+ \lambda_{23} (\eta^\dagger \eta)(\chi^\dagger \chi) \nonumber\\
    &+ \zeta_{12} (\rho^\dagger \eta)(\eta^\dagger \rho)+ \zeta_{13} (\rho^\dagger \chi)(\chi^\dagger \rho)+ \zeta_{23} (\eta^\dagger \chi)(\chi^\dagger \eta) \nonumber\\
    &+\sqrt{2} f \epsilon^{ijk} \rho_i \eta_j \chi_k\,,
\end{align}
where further terms with an odd power of $\eta$ and of $\chi$ are assumed to be forbidden via a discrete symmetry (see, e.g., \cite{Escalona:2025jla}), while the $(\chi^\dagger \eta)(\chi^\dagger \eta)$ term can be avoided via assigning for example different charges under a global $U(1)$.
We will assume the parameters of \eqref{eqn:331potential} to take values such as to lead to the vacuum of Eqs.~\eqref{eq:vchi} and \eqref{eq:ver} at low energies. After 8 Goldstone bosons have been absorbed by the massive gauge bosons of the model, there remain four neutral scalars $h,H,H',Y$, two charged scalars $H_1^\pm, H_2^\pm$, and a pseudoscalar $A$, in the physical scalar spectrum. 

The fermion sector of the 3-3-1RHN model is given by\footnote{Note that while anomaly cancellation requires the same amount of triplets and antitriplets under $SU(3)_L$ in the fermion spectrum, there is freedom to choose which quark generation transforms in the conjugate representation of the two others. Here, we choose to treat the third quark generation differently.}

\begin{equation}
\begin{split}
    L_L^i=  \begin{pmatrix}
       \nu_L \\
       e_L \\
       (\nu_R)^c
   \end{pmatrix} & \sim (1,3,-1/3)\,,\ \ell_R^i \sim (1,1,-1) \,,
\end{split}
\end{equation}
\begin{equation}
\begin{split}
   Q_L^\alpha=  \begin{pmatrix}
       d_L^\alpha \\
       - u_L^\alpha \\
       d_L^{\prime \alpha}
   \end{pmatrix} & \sim(3, \bar 3,0) \,,\ d_R^\alpha  \sim (3,1,-1/3) \,,\\[3mm]  u_R^\alpha & \sim (3,1,2/3) \,,\ d_R^{\prime\alpha} \sim (3,1,-1/3) \,,\\[3mm]
   Q_L^3=  \begin{pmatrix}
       u_L^3 \\
       d_L^3 \\
       u_L^{\prime 3}
   \end{pmatrix} & \sim(3,3,1/3) \,,\ u_R^3  \sim (3,1,2/3) \,,\\[3mm] d_R^3 & \sim (3,1,-1/3) \,,\ u_R^{\prime3} \sim (3,1,2/3)  \,,
\end{split}
\end{equation}
with $i=1,2,3$ and $\alpha=1,2$, and features six non-SM fermions. These are denoted by primed letters, with the latter indicating their EM charge, in analogy to the corresponding SM quarks.
All quarks and charged leptons obtain masses via the Yukawa interactions\footnote{For the purpose of this analysis, we neglect neutrino masses.}\cite{Montero:1992jk,Foot:1994ym}
\begin{equation}
\label{eq:Yuk3-3-1}
    \begin{split}
        {\cal L}_Y & = (y^\eta_d)_{\alpha j}\, \bar Q_L^\alpha\, \eta^\ast d_R^j 
        +(y^\eta_u)_{j}\, \bar Q_L^3\, \eta\, u_R^j \\
        & +(y^\rho_u)_{\alpha j}\, \bar Q_L^\alpha\, \rho^\ast u_R^j 
        +(y^\rho_d)_{j}\, \bar Q_L^3\, \rho\, d_R^j  + (y^\rho_\ell)_{ij}\, \bar L_L^i\, \rho\, \ell_R^j \\
        & + (y^\chi_d)_{\alpha\beta}\, \bar Q_L^\alpha\, \chi^\ast d_R^{\prime \beta} 
        +(y^\chi_u) \bar Q_L^3\, \chi\, u_R^{\prime 3} 
        + {\rm h.c.} \,,
    \end{split}
\end{equation}
where terms that would lead to mass mixings between SM-like fermions and the new fermions are avoided by extending the aforementioned discrete symmetry to the fermion sector~\cite{Escalona:2025jla}. From Eq.~\eqref{eq:Yuk3-3-1} it follows that the new fermions will naturally have masses of the order of $v_\chi \gg v_{\eta,\rho}$.

Given that our goal is merely to demonstrate an application of our method, we will adopt the decoupling limit from \cite{Escalona:2025jla} for simplicity, which corresponds to the off-diagonal mass matrix elements mixing the fluctuations around the vevs $(h_\rho,h_\eta)$ and $h_\chi$ being zero. The scalar mass matrix can then be diagonalised by a single angle $\varphi$, i.e.
\begin{align}
    \begin{pmatrix}
        h_\rho \\ h_\eta
    \end{pmatrix} = \begin{pmatrix}
        \cos\varphi & \sin\varphi\\
        -\sin\varphi & \cos\varphi
    \end{pmatrix} \begin{pmatrix}
        H \\ h
    \end{pmatrix} \,.
\end{align}
The 13 free parameters appearing in the potential \eqref{eqn:331potential} are thus fixed by two decoupling limit conditions, and the following 11 input observables: 
\begin{align}\label{eqn:331-input}
    m_h, v, \tilde\beta, \varphi+\tilde\beta, v_\chi, m_H, m_{H'}, m_{A}, m_{H_1^\pm}, m_{H_2^\pm}, m_Y.
\end{align}
Here, $v^2 = v_\rho^2+v_\eta^2$ and $\tilde\beta = \frac{v_\eta}{v_\rho}$ is chosen to be reminiscent of the 2HDM $\beta$ parameter.
\subsection{Vacuum Metastability}

The potential \eqref{eqn:331potential} itself is not biquadratic. However, it has been shown that by writing the fields in polar form
\begin{align}
    \Phi_i = \sqrt\frac{r_i}{2} e^{i\gamma_i}\begin{pmatrix}
        \sin(a_i)\cos(b_i)\\ e^{i\alpha_i}\sin(a_i)\sin(b_i) \\ e^{i\beta_i}\cos(a_i)
    \end{pmatrix},
\end{align}
where $\Phi_i = (\rho,\eta,\chi)_i$, and minimizing over the angular variables $\bm\theta =(a_i,b_i,\alpha_i,\beta_i,\gamma_i)$, we can write the quartic part of the potential to be biquadratic in $\sqrt{r_i}$ \cite{Costantini:2020xrn,Faro:2019vcd}\footnote{Note the difference in convention due to the inclusion of a factor $\sqrt{2}$. This is done to canonically normalize the fields so that we can write $r_i = v_i +h_i + i \eta_i$.}:
\begin{align}\label{eqn:331-biquadratic}
    \min_{\bm\theta} V^{\{4\}}_{331}(\Phi) = \min_{A} \sum_{ij} (\Lambda_{ij}^R+\Lambda_{ij}^A)r_i r_j.
\end{align}
Here, $A \in\{T_1,T_2, T_3, T, NT\}$ labels the various angular minima one can obtain by minimizing over $\bm\theta$. Also, for the purpose of minimization, $r$ is restricted to either the standard simplex ($\sum_i r_i=1,r_i>0$), or for $A=NT$, the triangular region $\Delta$ defined in \cite{Faro:2019vcd} which notably still has $\sum_i r_i = 1$. The $\Lambda$ are given by
\begin{align}\label{eqn:331matrices}
    \Lambda^R &= \frac{1}{8}\begin{pmatrix}
        2\lambda_1 & \lambda_{12}' & \lambda_{13}'\\
        \lambda_{12}' & 2\lambda_2 & \lambda_{23}'\\
        \lambda_{13}' & \lambda_{23}' & 2\lambda_3'
    \end{pmatrix},\nonumber\\
    \Lambda^{T_1}& = -\frac{1}{8}\begin{pmatrix}
        0 &0 & \zeta_{13}\\
        0 & 0 & \zeta_{23}\\
        \zeta_{13} & \zeta_{23} & 0
    \end{pmatrix}, 
    &\Lambda^{T_2}&= -\frac{1}{8}\begin{pmatrix}
        0 & \zeta_{12} &0\\
        \zeta_{12} & 0 & \zeta_{23}\\
        0 & \zeta_{23} & 0
    \end{pmatrix},\nonumber\\
    \Lambda^{T_3}&= -\frac{1}{8}\begin{pmatrix}
        0 & \zeta_{12} & \zeta_{13}\\
        \zeta_{12} & 0 & 0\\
        \zeta_{13} & 0 & 0
    \end{pmatrix}, 
    &\Lambda^{T}& = -\frac{1}{8}\begin{pmatrix}
        0 & \zeta_{12} & \zeta_{13}\\
        \zeta_{12} & 0 & \zeta_{23}\\
        \zeta_{13} & \zeta_{23} & 0
    \end{pmatrix},\nonumber\\
    \Lambda^{NT} & = \mathrlap{-\frac{1}{16}\begin{pmatrix}
        \frac{\zeta_{12}\zeta_{13}}{\zeta_{23}} & \zeta_{12} & \zeta_{13}\\
        \zeta_{12} & \frac{\zeta_{12}\zeta_{23}}{\zeta_{13}} & \zeta_{23}\\
        \zeta_{13} & \zeta_{23} & \frac{\zeta_{13}\zeta_{23}}{\zeta_{12}}
    \end{pmatrix}\,.}
\end{align}
Note that for $A=NT$, the condition $r\in\Delta$ does not match the constraint we imposed in \eqref{eqn:constrainedproblem}. So, our solutions in Table~\ref{tab:lambdaeff} do not directly apply. We thus define the following transformation, similar to \cite{Faro:2019vcd},
\begin{align}
    R &=\begin{pmatrix}
        0& \frac{|\zeta_{23}|}{|\zeta_{23}|+|\zeta_{12}|} & \frac{|\zeta_{23}|}{|\zeta_{23}|+|\zeta_{13}|}\\
        \frac{|\zeta_{13}|}{|\zeta_{13}|+|\zeta_{12}|} & 0 & \frac{|\zeta_{13}|}{|\zeta_{13}|+|\zeta_{23}|}\\
        \frac{|\zeta_{12}|}{|\zeta_{13}|+|\zeta_{12}|} & \frac{|\zeta_{12}|}{|\zeta_{23}|+|\zeta_{12}|} & 0
    \end{pmatrix}.
\end{align}
This matrix converts the constrained minimization problem to be equivalent to \eqref{eqn:constrainedproblem} \cite{Faro:2019vcd}, keeping the normalization condition invariant 
\begin{align}
    \sum_i r_i  =1 \iff \sum x_i := \sum_{ij} R_{ij}^{-1}r_j = 1.
\end{align}
Thus, we need only minimize the biquadratic form $R^T(\Lambda^R+\Lambda^{NT}) R$ over the standard simplex. 
As all the matrices are again $3\times 3$, we can easily iterate through all the submatrices and use our expressions for $\lambda_\text{eff}$ in Table~\ref{tab:lambdaeff}, subject to the condition $r_i, x_i >0$ for all $i$.

\subsection{Theoretical Constraints}

As in the 2HDM+a example, we consider all P, PU, and BFB constraints on the 3-3-1 model. For perturbativity, we simply require all $\lambda_i,\lambda_{ij}, \zeta_{ij}<4\pi$. For PU, we use the results from \cite{Costantini:2020xrn}, with which we share conventions for the form of the potential \eqref{eqn:331potential}. For BFB, we follow the same procedure as in the 2HDM+a model, and impose the implication $\bm{u}_K >0 \implies \lambda^K_\text{eff}>0$ at tree level for each of the $r_i,x_i$ in all the angular minima $A$.

Note that due to the mass hierarchy $v,m_h \ll v_\chi, m_H$, the quartic parameters in the potential are very sensitive to our choice of input parameters \eqref{eqn:331-input}. In particular, denoting $\varepsilon=v/v_\chi$, and keeping the quartics perturbatively small $\lambda_i,\lambda_{ij},\zeta_{ij} = O(\varepsilon^0)$, we are forced to impose
\begin{align}\label{eqn:331-perturbativity}
    \frac{m_{H_2^\pm}-m_H}{v_\chi},\, \frac{m_{A}-m_H}{v_\chi} ,\, \cos( \varphi+\tilde\beta) = O(\varepsilon^2).
\end{align}

\begin{figure*}[t]
		\resizebox{1\textwidth}{!}{
			\includegraphics[height=0.3\textwidth]{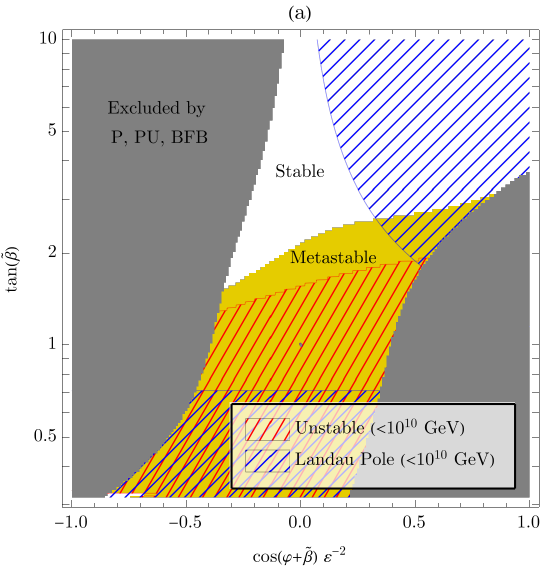}
			\includegraphics[height=0.3\textwidth]{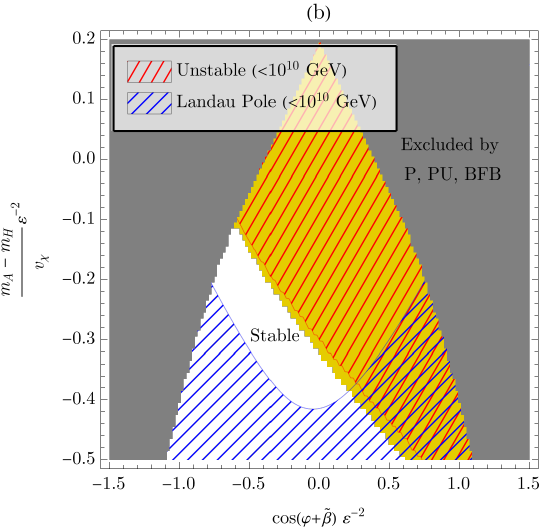}
		}
		\caption{Parameter scans of the 3-3-1 model phase diagram (a) in the $\cos({ \varphi }+\tilde\beta) \times \tan\tilde\beta$ plane and (b) in the $ \cos(\varphi+\tilde\beta) \times m_A$ plane. All other parameters, including $m_H$ and $v_\chi$, are fixed in \eqref{eqn:331-paramchoices}, while $\varepsilon$ is defined before \eqref{eqn:331-perturbativity} and is approximately $0.018$ here.}
		\label{fig:scan-331}
\end{figure*}

\subsection{Experimental Constraints}

As with the 2HDM+a model, we will only consider a limited set of experimental constraints. In particular, based on dilepton data from the LHC, Alves at al. \cite{Alves:2022hcp} obtained a bound on $m_{Z'}>4\,$TeV, with projected bounds at $m_{Z'}\gtrsim 10\,$TeV for future colliders. This translates to a current bound of $v_{\chi}\gtrsim 10\,$TeV, and projected bounds of the order of $v_\chi \gtrsim 20\,$TeV. As such, we will set $v_\chi =20\,$TeV to avoid these constraints.  
We also consider the FCNC results provided in \cite{Escalona:2025jla}. However, we find that for $m_H=10\,$TeV and $v_\chi = 20\,$TeV, the constraints \eqref{eqn:331-perturbativity} are far more stringent.

Finally, to make contact with physics at the electroweak scale, we make use of SM values from \cite{ParticleDataGroup:2024cfk},
\begin{align}
    \lambda(m_h) &= 0.260  & m_h(m_h)&=125\,\text{GeV}\nonumber  \\
    m_W&=80.4\,\text{GeV} & m_Z &= 91.2\,\text{GeV} \nonumber  \\
     y_\tau(m_\tau) &= 0.0096& y_b(m_b) &= 0.023\nonumber \\
     y_t(m_t)&=0.933&  g_1(m_Z) &= 0.357\nonumber  \\
   g_2(m_Z)&=0.652& g_3(m_Z) &= 1.22   
\end{align}
and run them up using SARAH \cite{Staub:2015kfa} two-loop RGEs to the 3-3-1 scale $\mu_{331} = 10\,$TeV. Note that we use SM RGEs here, since we expect all beyond-SM particles to generically have masses of the order of $\sim v_\chi$. Also $m_h(\mu)$ here is interpreted not as the physical pole mass, but as the tree-level mass of the theory, and thus depends on the renormalization scale. We thus obtain the following inputs for the 3-3-1 model:
\begin{align}\label{eqn:331-expinputs}
    m_h(\mu_{331}) &= 132\,\mathrm{GeV},\ &v(\mu_{331})&=364\,\mathrm{GeV},\nonumber\\ 
    g_X(\mu_{331}) &= 0.390, \ & g(\mu_{331})&=0.628,\hfill\nonumber\\ 
    (y_\ell^\rho)_{33}(\mu_{331}) &= \frac{0.011}{\cos\tilde\beta},\ & (y_d^\rho)_3(\mu_{331}) &= \frac{0.013}{\cos\tilde\beta}, \nonumber\\ 
    (y_u^\eta)_3(\mu_{331})&=\frac{0.7729}{\sin\tilde\beta}.
\end{align}
Note that the yukawa couplings of the first two generations of SM fermions are set to zero as they have negligible impact on vacuum decay constraints. Here, $m_h$ and $v$ are also interpreted as tree-level inputs to the 3-3-1 model \eqref{eqn:331-input}. Indeed, writing the SM Higgs as $h$, we find that power counting according to \eqref{eqn:331-perturbativity} gives a tree-level $hh\to hh$ scattering amplitude of (up to factors of $i$),
\begin{align}
    \frac{\partial^4 V_{331}}{\partial h^4} =\frac{ 3m_h^2}{v^2}+ O\left(\varepsilon^2\right)
\end{align}
which recovers the SM result. To match the 3-3-1 model to the SM, we also require the tree level Higgs mass $m_h$ to be equal in the two theories. So, taken together, we can to leading order match the SM and 3-3-1 quantities at the 3-3-1 scale with $m_h^\text{SM}(\mu_{331}) = m_h^\text{331}(\mu_{331})$ and $v^\text{SM}(\mu_{331})=v^\text{331}(\mu_{331})$. 

\subsection{Results}

Unlike in the 2HDM+a model, where we set the upper bound $\Lambda = 10^{10}\,$GeV, for the 3-3-1 model we set the arbitrary upper bound to $\Lambda=10^{12}\,$GeV. This reflects our assumption that the 3-3-1 is valid at $\mu_{331}$ which is approximately two orders of magnitude above the electroweak scale. Thus, the 2HDM+a and 3-3-1 bounds in a rough sense reflect a similar confidence in exclusion. 

To allow for a straightforward comparison with flavor constraints in \cite{Escalona:2025jla}, we scan through the $\cos(\varphi+\tilde\beta)$ versus $\tan\tilde\beta$ plane in Fig.~\sref{fig:scan-331}{(a)}. In Fig.\,\sref{fig:scan-331}{(b)} we also present one of the more nontrivial bounds we found, in the $ \cos(\varphi +\tilde\beta) \times m_A$ plane. In both of these cases, for the parameters kept constant, we have chosen:
\begin{align}\label{eqn:331-paramchoices}
    m_{H}=m_{H'}=m_{H_1^\pm}=m_{H_2^\pm}=m_A=m_Y&=10\,\text{TeV},\nonumber\\
    v_\chi = 20\,\text{TeV}\comma \tan\tilde\beta&=1\comma \nonumber\\
    (y_d^\chi)_{11}=(y_d^\chi)_{22}=y_u^\chi&=0.5\,,
\end{align}
where off diagonal yukawas are set to zero. The rationale behind the choice of yukawa couplings is that we want to keep the exotic quark masses of the order of $v_\chi$, while avoiding the strict vacuum decay constraints at large yukawa couplings. Indeed, we find that similar to the 2HDM+a model (see e.g. Fig.\,\ref{fig:scan-2hdma-yx-l}), large exotic yukawa couplings of order $y^\chi\sim0.8$ destabilize the vacuum strongly. However, the plots of $m_H$ versus $y^\chi$, where $m_H$ indirectly controls the quartic parameters in the potential~\eqref{eqn:331potential}, look qualitatively similar to the plots in Fig.\,\ref{fig:scan-2hdma-yx-l}. Thus, we will not show them here to avoid redundancy.

It is interesting to note that the excluded regions in the $\cos(\varphi+\tilde\beta)$ versus $\tan\tilde\beta$ plane look very different from the corresponding ones of the 2HDM+a case. Indeed, we find that for the 3-3-1 model, the low $\tan\tilde\beta \lesssim 1.5$ range is excluded by vacuum decay for the full range of scalar mixings $\varphi$, at least for our benchmark choices \eqref{eqn:331-paramchoices}. However, from Fig.~\sref{fig:scan-331}{(b)}, we see that small pseudoscalar mass splittings of order $m_A-m_H\approx -2\,$GeV can recover absolute stability near $\tan {\tilde \beta}=1$, so the bound of Fig.~\sref{fig:scan-331}{(a)} is not absolute. The allowed gap in Fig.~\sref{fig:scan-331}{(b)} does close though, at around $\tan{\tilde \beta} \approx 0.75$.

\begin{figure}[h!]
		\includegraphics[width=\linewidth]{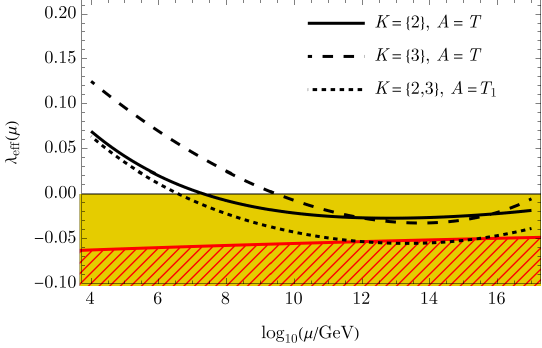}
		\caption{Running of $\lambda_\text{eff}^K$ for choices of $K$ and biquadratic forms in \eqref{eqn:331-biquadratic}. $K$ and $A$ are defined as in \eqref{eqn:completelambdaeff} and \eqref{eqn:331-biquadratic} respectively. Only the $\lambda_\text{eff}^K$ running below 0 have been plotted for comparison. The red hatched region corresponds to values of $\lambda_\text{eff}$ that lead to an unstable vacuum as per \eqref{eqn:lambdabound}.}
		\label{fig:lambdaeff-3-3-1}
\end{figure}

Finally, to demonstrate the necessity of including the bounces in the nontrivial directions, we plot an example of the running of the various $\lambda_\text{eff}$ in Fig.\,\ref{fig:lambdaeff-3-3-1}. Here, we set
\begin{align}
    m_{H}=m_{H'}=m_{H_1^\pm}=m_{H_2^\pm}=m_A=10\,\text{TeV}\nonumber\\
    m_Y=5\,\text{TeV}\comma v_\chi = 20\,\text{TeV} \nonumber\\
    \cos({ \varphi}+\tilde\beta)=0 \comma \tan\tilde\beta=2.5\comma \nonumber\\
    (y_d^\chi)_{11}=(y_d^\chi)_{22}=0.5\comma y_u^\chi=0.87.
\end{align}
Note that only $y^\chi_u$ and $m_Y$ have been tuned to realize the depicted scenario. We see in Fig.\,\ref{fig:lambdaeff-3-3-1} that the bounces corresponding to only a single field direction $K=\{2\}$ and $K=\{3\}$ predict a metastable vacuum, while the nontrivial field direction given by $K=\{2,3\}$ and $A=T_1$ predicts an unstable vacuum. All the bounce solutions considered here are thus necessary for a complete characterization of vacuum decay constraints in the 3-3-1 model at leading order.

\section{Conclusion}
\label{Sec:conclusion}

In this paper, we developed a general framework to compute the exact leading order vacuum decay rates in multi-scalar theories, in the high energy limit where radiative corrections dominate, $\phi \to \infty$. We demonstrated that, in this limit, the exponentially dominant bounces are constrained to move only in a radial line. This thus generalizes the one-dimensional solution \eqref{eqn:onedimensionalbounceaction} to apply to the multi-field case as in equation \eqref{eqn:bounceactioneff}. Furthermore, we showed that, when the scalar sector can be represented as a biquadratic form, the general optimization problem can be reduced to solving a series of linear equations, which can be performed efficiently using existing linear algebra libraries such as LAPACK~\cite{laug}. The remaining bottleneck in this approach is the number of subsets $K$ of fields we need to consider in \eqref{eqn:completelambdaeff}. This can be optimized by exploiting symmetries of the scalar sector, as we did for the 2HDM+a model in section \ref{Sec:2HDMa} and the 3-3-1 model in section \ref{Sec:3-3-1}. In both of these models, the symmetries were strong enough that we could produce useful analytical forms for the bounce action by employing the general results we provided for multi-scalar theories in Table~\ref{tab:lambdaeff} and equations \eqref{eqn:2hdmamatrices} and \eqref{eqn:331matrices}. 

Next, we applied our methods to obtain parameter-space scans for vacuum decay constraints on the 2HDM+a and 3-3-1 models. In both cases we found that, in phenomenologically relevant regions, high-scale validity typically dominates existing experimental bounds. In particular, we found that vacuum stability typically favours regions with yukawa couplings $\lesssim 1$. In a scenario where the top quark couples to only one of many vevs, this vev is then constrained to be of $\mathcal{O}(246\,\text{GeV}),$ lest the top yukawa coupling $y_t\propto m_t/v$ becomes too large. Finally, as illustrated in Fig.\,\ref{fig:lambdaeff-3-3-1}, a complete analysis of vacuum decay constraints in multi-scalar theories at leading order does require us to include all solutions in Eqs.~\eqref{eqn:lambdaeff} or \eqref{eqn:completelambdaeff}. Neglecting multi-field solutions with $\|K\|>1$ can lead to unstable points being mischaracterized as stable.

\section*{Acknowledgements}
FG acknowledges useful discussions with Farinaldo S. Queiroz. RW thanks Cedric Simenel for helpful discussions. GB was supported by the Australian Government through the Australian Research Council Centre of Excellence for Dark Matter Particle Physics (CDM, CE200100008). 
NK was supported by an Australian Government Research Training Program Scholarship.
RW was supported by the Australian National University through the Dunbar Physics Honours Scholarship.

\bibliographystyle{utphys}    %
\bibliography{refs}     %

\end{document}